\documentclass[paper=letter,abstracton,11pt]{scrartcl}

\newcommand{\fullversiononly}[1]{#1}
\newcommand{\eabstronly}[1]{}

\usepackage{fullpage}
\usepackage[short]{privbib}
\usepackage{privmath}
\usepackage{enumerate,cite}
\usepackage{epsf,here,amssymb,bbm,amsmath,exscale}
\usepackage[font=small,format=plain,labelfont=bf,up,textfont=rm,up]{caption}
\usepackage{hyperref}

\usepackage[textsize=tiny,disable]{todonotes}
\newcommand{\pwtodo}[1]{\todo[color=blue!30]{#1}}
\newcommand{\mdtodo}[1]{\todo[color=green!30]{#1}}
\newcommand{\mdignore}[1]{}

\newcommand{\St}{{\cal S}}

\newcommand{\ie}{i.\,e.}
\newcommand{\eg}{e.\,g.}

\newcommand{\sln}[1]{{\ln^{2}(#1)}}
\newcommand{\mdist}[1]{\left\|{#1}\right\|_1}

\newcommand\nat{{\mathbb N}}
\newcommand\integer{{\mathbb Z}}

\newcommand\real{{\mathbb R}}
\newcommand\IN{\nat}
\newcommand\IZ{\integer}

\newcommand\IR{\real}

\newcommand\ZD{{\integer^D}}

\newtheorem{theorem}{Theorem}
\newtheorem{claim}[theorem]{Claim}
\newtheorem{lemma}[theorem]{Lemma}

\newtheorem{remark}{Remark}

\title{Tight Lower Bounds for Greedy Routing in Higher-Dimensional Small-World Grids}
\author{
   \Large Martin Dietzfelbinger\\
   \normalsize Faculty of Computer Science and Automation\\
   \normalsize  Ilmenau University of Technology\\
   \normalsize  98694 Ilmenau, Germany\\
   \normalsize  \texttt{martin.dietzfelbinger@tu-ilmenau.de}\\
   \and
   \Large  Philipp Woelfel\\
   \normalsize  Department of Computer Science\\
   \normalsize  University of Calgary\\
   \normalsize  Calgary, AB T3A1R9, Canada\\
   \normalsize  \texttt{woelfel@ucalgary.ca}\\
}

\begin{document}
\maketitle
\begin{abstract}
  We consider Kleinberg's celebrated small world graph model \cite{Kle2000b,Kle2000a}, in which a $D$-dimensional grid $\{0,\dots,n-1\}^D$ is augmented with a constant number of additional unidirectional edges leaving each node.
  These \emph{long range} edges are determined at random according to a probability distribution (the augmenting distribution), which is the same for each node.
  Kleinberg suggested using the \emph{inverse $D$-th power distribution}, in which node $v$ is the long range contact of node $u$ with a probability proportional to ${\mdist{u-v}}^{-D}$.
  He showed that such an augmenting distribution allows to route a message efficiently in the resulting random graph:
  The greedy algorithm, where in each intermediate node the message travels over a link that brings the message closest to the target w.r.t. the Manhattan distance, finds a path of expected length $O\bigl((\log n)^2\bigr)$ between any two nodes.
  In this paper we prove that greedy routing does not perform asymptotically better for any uniform and isotropic augmenting distribution, \ie, the probability that node $u$ has a particular long range contact $v$ is independent of the labels of $u$ and $v$ and only a function of $\mdist{u-v}$.
  
  In order to obtain the result, we introduce a novel proof technique: We define a so-called \emph{budget game}, in which a token travels over a game board, from one end to the other, while the player manages a ``probability budget''.
  In each round, the player ``bets'' part of her remaining probability budget on step sizes.
  A step size is chosen at random according to a probability distribution of the player's bet.
  The token then makes progress as determined by the chosen step size, while some of the player's bet is removed from her probability budget.
  We prove a tight lower bound for such a budget game, and then obtain a lower bound for greedy routing in the $D$-dimensional grid by a reduction.  
\end{abstract}
\newpage

\section{Introduction}\label{sect:introduction}
In 2000, Kleinberg proposed his celebrated \emph{small world graph model} to study the performance of decentralized search in augmented networks \cite{Kle2000b,Kle2000a}.
His original network model was based on an $n\times n$ grid with vertex set $\bigl\{(x,y):x,y\in \{0,\dots,n-1\}\bigr\}$, and the Manhattan (or lattice) distance as a distance metric. (I.\,e., the distance between two points $u=(x,y)$ and $u'=(x',y')$ is $\mdist{u-u'}=|x-x'|+|y-y'|$.)
For some constant $p$, every node $u$ has a directed edge to every other node $v$ within distance $p$ ($v$ is a \emph{local contact} of $u$).
In addition, for some universal constants $r,q$, every node $u$ has $q$ additional directed edges to other nodes, called \emph{long range contacts}.
The long range contacts of $u$ are determined by $q$ independent random trials, where a particular node $v$ is chosen according to the \emph{inverse $r$-th power distribution}, \ie, with probability proportional to ${\mdist{u,v}}^{-r}$.
Kleinberg showed that in such a network with $r=2$ and $p=q=1$ a simple \emph{decentralized} routing algorithm can efficiently route a message from any given source to any given target.
More precisely, he proposed the greedy algorithm, where in each intermediate node the message travels over a link whose endpoint minimizes the remaining Manhattan distance to the target.
Kleinberg showed that this algorithm yields a route of expected length $O\bigl((\log n)^2\bigr)$.
This result can be generalized to $D$ dimensions for any constant $D\geq 1$:
In this case, the expected greedy routing length is quadratic in $\log n$ if $r=D$.
The parameter $r=D$ is in fact optimal, as for $r\neq D$ no decentralized algorithm can find a route in poly-logarithmic time (see also \cite{Kle2001a}).

Following Kleinberg's seminal work, more general models were studied, and over the last decade a rich theory of decentralized algorithms and their performance in augmented networks was developed.
(See for example the surveys by Fraigniaud \cite{Fra2007a} and Kleinberg \cite{Kle2006a}.)
An augmented network, as defined in its general form for example by Fraigniaud, Lebhar, and Lotker~\cite{FLL2006a}, consists of a directed base graph of $N$ nodes
which defines the local contacts of nodes, together with a probability distribution for long range contacts, called \emph{augmenting distribution}.
An instance of an augmented network is a random graph which has all the vertices and directed edges from the base graph as well as additional directed long range edges from each node to its long range contacts, which are chosen at random according to the augmenting distribution.
Usually, a distance metric is associated with the network.
The distance between two nodes can be, for example, the length of the shortest path between them on the base graph.
For algorithmic purposes, the distance between nodes is expected to be easy to compute or to approximate locally \cite{Fra2007a}.
The augmenting distribution is usually \emph{uniform} and \emph{isotropic} in the sense that the probability that node $u$ has a particular long range contact $v$ does not depend on the labels of $u$ and $v$, but only on the distance from $u$ to $v$.

Uniform isotropic augmented networks were first considered by Watts and Strogatz~\cite{WS1998a} as a mathematical model to understand the ``small-world phenomenon'' occurring in social networks and the web.
The most prominent example of the phenomenon is the existence of short chains of acquaintances, or ``six degrees of separation'', between individuals in the United States, as observed in Milgram's famous experiment \cite{Mil1967a}.

While earlier work, including that of Watts and Strogatz's, was concerned with structural properties of augmented networks,
Kleinberg~\cite{Kle2000b,Kle2000a} started to study the question of \emph{navigability}, \ie, whether short paths can be found efficiently in augmented networks.
He suggested using \emph{decentralized} algorithms for routing messages.
In such algorithms routing decisions are made locally: 
The node that holds the message determines the (local or long range) contact to send the message to, using only local information.
The message holder can use knowledge about the base graph, the location of the target on the base graph, and the locations and long-range contacts of nodes that have come in contact with the messages.
However, the message holder has no information about the long range contacts of nodes that have not yet touched the message.

The simplest, most natural and consequently best studied decentralized algorithm is the greedy algorithm, in which the message holder passes the message on to its local or long range contact which is closest to the target (according to the distance metric associated with the base graph).
This led to the definition of the \emph{greedy diameter} of an augmented graph, which is the maximum of the expected routing times of all pairs of source and target nodes in the base graph \cite{FLL2006a}.
Aspnes, Diamadi, and Shah \cite{ADS2002a} observed the usefulness of greedy routing in augmented graphs 
for peer-to-peer networks; 
several peer-to-peer systems such as Chord~\cite{Gane2004a,Sto2003a}, Symphony~\cite{MBR2003a} or Randomized-Chord~\cite{Gumm2003a,Xu2003a} are based on augmented rings.
In all those systems the augmenting distribution is based on the inverse $r$-th power distribution, and the expected greedy routing time between any pair of nodes is $O\bigl((\log n)^2\bigr)$ (or $O\bigl((\log n)^2/k\bigr)$ if the number of long range contacts per node, $k$, is super-constant but logarithmic).

\pwtodo{Small modifications to this paragraph.}
This raised the question whether there exist uniform and isotropic augmenting distributions that yield faster greedy routing times.
Several attempts were made to find an answer to this question, but until now succeeded only in restricted settings.
In fact, general lower bounds obtained in the past hold only for the one-dimensional case, where the base graph is the line or the ring:
Flammini, Moscar\-del\-li, Navarra, and P{\'e}rennes~\cite{FMNP2005a} showed an optimal lower bound of $\Omega((\log n)^2)$ for the ``oblivious diameter'' of the augmented line with an augmenting distribution that satisfies a monotonicity requirement.
(The oblivious diameter is a measure that is potentially larger than the greedy diameter.)
A tight lower bound for the greedy diameter of the augmented ring, where the augmenting distribution is the $r$-th power distribution with $r=1$, was obtained by Barri\`{e}re, Fraigniaud, Kranakis, and Krizanc~\cite{Barr2001a}.
Almost tight lower bounds for all uniform and isotropic augmenting distributions were provided by Aspnes, Diamadi, Shah~\cite{ADS2002a} and were subsequently improved by Giakkoupis and Hadzilacos~\cite{GH2007a}.
Tight lower bounds for the greedy diameter of the line of length $n$ with one long range contact per node were obtained by Dietzfelbinger, Rowe, Wegener, and Woelfel~\cite{DRWW2010a}.
Finally, Dietzfelbinger and Woelfel~\cite{DW2009a} proved a tight lower bound of $\Omega\bigl((\log n)^2/k\bigr)$ for the greedy routing time between randomly chosen start and target nodes for the augmented ring and line with $k=O(\log n)$ long range contacts per node.
(The lower bound also holds if the source and target nodes are chosen at random.)

\subsection{Results}
In contrast to the one-dimensional case, no very general lower bounds were known for the two-dimensional or even higher dimensional case.
Kleinberg~\cite{Kle2000a} proved a lower bound of $n^{\Omega(1)}$ for the expected routing time of any decentralized algorithm in the augmented two-dimensional grid, if the augmenting distribution is the inverse $r$-th power distribution for $r\neq 2$.
The result extends to the $D$-dimensional case if $r\neq D$.
In the $D$-dimensional grid augmented with the inverse $D$-th power distribution, the greedy diameter is $\Theta\bigl((\log n)^2\bigr)$, \ie, Kleinberg's upper bound is tight \cite{FGP2006a,MN2004a}.
Similar lower bounds hold also if the $D$-dimensional torus (\ie, the grid with wrap-around links) is used as a base graph.

We are not aware of any lower bounds for general uniform and isotropic augmenting distributions in the $D$-dimensional grid or torus for $D\geq 2$.
This raises the question whether for $D\geq 2$ the inverse $D$-th power distribution is in fact optimal among all uniform and isotropic augmenting distributions.
In this paper we answer the question in the affirmative.
More precisely, we consider as a base graph the $D$-dimensional torus $\{0,\dots,n-1\}^D$ with $N=n^D$ nodes.
We show for any such augmenting distribution and any two nodes $s,t$ with $\mdist{s-t}\leq n/4$ that the expected greedy routing time from $s$ to $t$ is $\Omega\bigl((\log \mdist{s-t})^2\bigr)$.
The same is true for the $D$-dimensional grid as a base graph, as long as $s$ and $t$ have distance at least $2\mdist{s-t}$ from the border of the grid.

Note that the restriction of isotropy is necessary for such a strong lower bound: It prevents us
from defining for some given pair $(s,t)$ an augmenting distribution, where there is always a short greedy route from $s$ to $t$ (for example, because $s$ has long range contact $t$ with probability 1).
On the other hand, non-isotropic but uniform augmenting distributions might still have a large greedy diameter.

For simplicity of exposition, we consider only the case that every node has one long range contact.
It is not hard to see that our lower bound extends to the case where every node has a constant number of long range contacts chosen independently at random according to the same augmenting distribution.
Similarly, the lower bound result is not affected if in the base graph any two nodes within Manhattan distance $p$ are connected, where $p$ is some constant.

Note that the lower bound holds for any source/target pair in the higher dimensional torus or grid.
This is not the case for $D=1$ (where the base graph is a ring or a line): For example the augmenting distribution may have the property that each node has a long range contact at distance $n/2$ with high probability.
If this is the case, the expected greedy routing time can be very small if the distance between the source and target is exactly $n/2$.
In higher dimensions this cannot happen, as for a given distance $d$ and a given node $u$ there are sufficiently many choices of possible long range contacts $v$ with $\mdist{u-v}=d$.

\subsection{The Lower Bound Technique}
\pwtodo{Two new paragraphs. Please check. Perhaps move this somewhere elsewhere?}
\mdtodo{Position is good. Little corrections.}
The lower bound proofs for the one-dimensional case were of increasing difficulty.
The result by Dietzfelbinger~\emph{et al.}~\cite{DRWW2010a} uses a quite complicated potential function.
The lower bound in~\cite{DW2009a} is based on an argument that with each step made by the greedy routing algorithm associates a \emph{cost} and shows that the expected cost per step is constant, while the total cost of an entire run is $\Omega(\log^2 n)$.
Both the potential function of \cite{DRWW2010a} and the cost function of \cite{DW2009a} are rather difficult to motivate and lead to verifiable but not very intuitive proofs.

The proof presented in this paper is quite different from any previous lower bound techniques we know of, and in particular from those used in the one-dimensional case~\cite{ADS2002a,GH2007a,DRWW2010a,DW2009a}.
We also believe that it is significantly more intuitive, and, while technically challenging, the general approach can be motivated more easily.

\pwtodo{I changed the following paragraph. Are the last two sentences too much?}
\mdtodo{Changed again.}
We define a probabilistic game, called \emph{budget game}, played in rounds by a single player.
Then we prove a lower bound on the expected number of rounds it takes the player to finish the game.
Finally we show that the expected number of rounds is a lower bound for the expected routing time in the $D$-dimensional torus.
The lower bound for the budget game is proved by induction.
The inductive hypothesis can be motivated easily. The inductive step is somewhat involved, but not too hard to verify.
The biggest problem to overcome in the course of building the overall argument was in designing the game,
tuning its details, and choosing its parameters in such a way that both the proof 
of the lower bound and the proof of the connection with the routing time goes through.

The budget game is played on a game board with $d_0+1$ spaces, labeled $d_0,d_0-1,\dots,0$ from left to right.
The player starts with a token at the leftmost space, $d_0$, and her goal is to move the token to space 0 in as few rounds as possible.
The player also manages a \emph{probability budget}.
In each round, the player ``bets'' a probability distribution $\beta$ over $\{1,\dots,d\}$, where $d$ is the current location of the token.
The total bet on step sizes larger than 1 must not exceed her remaining probability budget, $B$, \ie, $\beta\bigl([2,d]\bigr)\leq B$.
Moreover, in order to prevent the token from moving too fast, bets on step sizes larger than $d/2$ are also constrained.
Now a step size $i$ is chosen at random according to $\beta$, and the player can move her token $i$ positions to the right.
The player loses a portion of her probability budget, which is determined by how much she bet on step sizes larger than $i/2$ and by the amount of progress made 
(relative to the remaining distance).

The connection to greedy routing on a grid is as follows:
The remaining distance the player has to travel on the game board corresponds to the distance on the grid between the message holder, $u$, and the target.
The remaining probability budget, $B$, corresponds to the probability that $u$ has a long range contact which is closer to the target than $u$ itself.
The bet $\beta$ of a player is determined by the augmenting distribution: $\beta(i)$ is roughly the probability that $u$ has a long range contact $v$ whose distance to the target is by $i$ smaller than $u$'s distance to the target.
As the message gets closer to its target, longer distances of long range contacts become useless.
The probability of choosing long range contacts which are ``too far away'' is captured in the budget game by the budget loss.

The constraint in the budget game on bets on large step sizes, \ie, step sizes larger than $d/2$, is justified by the isotropy of the augmenting distribution and the topology of the base graph:
Suppose on the $D$-dimensional torus the message holder, $u$, is in distance $d$ of the target node.
Even if the augmenting distribution dictates that the randomly chosen long range contact $v$ of $u$ is with probability 1 in the ``optimal'' distance from node $u$ (whatever distance is optimal), the probability that $v$ will not be closer than $d/4$ to the target is constant.
This is because all nodes at the same distance from $u$ will be chosen with the same probability as long range contacts and only a constant fraction of them can be closer than $d/4$ to the target.
More generally, $u$'s long range contact will be closer than $d'\leq d/2$ to the target only with probability $O(d'/d)$.

There is quite some flexibility in defining the probability budget game precisely, \ie, how big the budget loss is in each step and how to constrain bets on step sizes larger than $d/2$.
We chose those parameters in such a way that it fits our application, but we believe that the general concept of a budget game is interesting in its own right.

In Section~\ref{sect:budget:game}, we describe the budget game in more detail and analyze it.
In Section~\ref{sect:greedy:routing} we prove the lower bound for the greedy routing time 
in augmented grids by showing that 
each distribution on the long range contacts gives rise to a 
strategy for the budget game, so that routing times translate into upper bounds 
on the expected time for the budget game. 
So our lower bound for the budget game leads to a lower bound for the routing situation. 

Throughout the text, ``$\ln$'' denotes the natural logarithm (to base $e$) and ``$\log$'' denotes the logarithm to base 2.
For ease of notation we write $\log^c x$ instead of $(\log x)^c$ and $\ln^c x$ instead of $(\ln x)^c$.

\section{The Budget Game}\label{sect:budget:game}

The budget game is played in rounds on the set $\{0,1,\ldots,d_0\}$ as ``gameboard''.
There are two parameters: $d$ is the remaining distance to be covered, and $B$ is the remaining \emph{budget}.
The ``player'' starts with a token at distance $d=d_0$ and wishes it to reach $0$. 
She manages a budget, which initially has value $B=B_0$. 
In a round, within the limits of the current budget and obeying some other restrictions 
the player can choose (``bet'') probabilities for certain distances to appear.
A step size $i$ is chosen at random, according to these probabilities. 
Then the distance $d$ is reduced by the chosen value $i$,
and the budget $B$ is reduced by an amount that depends on the jump distance $i$
and on the probabilities used in the betting. 

Our goal is to establish lower bounds on the expected number of moves needed
when starting in $d_0$ with budget $B_0$.

More precisely, when in state $(d,B)$, one round of the game comprises the following actions.
\begin{itemize}
  \item[1.] The player chooses a \emph{valid bet for budget $B$ and distance $d$}, which is a probability distribution $\beta$ over step sizes $\{1,\dots,d\}$, satisfying
  the following constraints:
  \begin{enumerate}
  \item[(B1)] $\beta([2,d])\leq B$;
  \item[(B2)]for all $1\leq j\leq d^* = \lceil d/2 \rceil = d-\lfloor d/2 \rfloor$: $$\displaystyle\frac{\beta\bigl([d-j+1,d]\bigr)}{\beta\bigl([d-d^*+1,d]\bigr)}\leq\frac{2j}{d^*}.$$
  \end{enumerate}
  The first constraint bounds the probability of choosing step sizes larger than 1 by the remaining probability budget.
  The second constraint bounds the conditional probability of choosing a very large step size 
  (close to the remaining distance $d$) given that the step size is at least half the remaining distance $d$.
  Note that the probabilities for step sizes up to 
   $d-d^*=\lfloor d/2\rfloor$ are not constrained; 
   for example, a strategy in which a step size of $\lfloor d/2\rfloor$ occurs with probability 1
    is allowed. 
    (Such a strategy would not be very efficient, though, as the probability 
    budget would get reduced by a constant in one step.)
  \item[2.] A random step size $i$ is chosen from $\{1,\ldots,d\}$ according to the probability distribution $\beta$. 
    \item[3.] Progress: Let $d=d-i$.
    \item[4.] Budget loss: If $i\geq 2$, let $B=B{-}(i/d)^{3/2}\beta\bigl((i/2,d]\bigr)$.%
    \footnote{For real numbers $x<y$
		the interval $(x,y]$ is $\{i\in\IN \mid x < i\le y\}$;
    similarly for closed intervals $[x,y]$.} 
    \quad(For $i=1$ the budget loss is 0.)\\
This budget loss depends on the ``order of magnitude'' 
of the jump $i$ actually chosen. 
The amount the player loses 
is determined by the total amount bet on the jump sizes in this order of magnitude and larger. 
If $i$ is close to $d$, a constant fraction of this is lost;
if $i$ is smaller than $d$ by a factor of $q$, 
   the loss is scaled down by a factor of $q^{3/2}$.
The basic idea would be to collect a fraction of $q$ of the amount bet,
to reflect the idea that $1/q$ jumps with a distance of about $d/i$
would suffice to reach the goal, so that no budget for jumps of this order
of magnitude is needed anymore. 
For technical reasons the formula is modified a little.
\end{itemize}
The game ends when $d$ becomes 0. 

Note that there is no budget loss in case the step size $i$ is 1,
but the player makes progress in this case, too.
Hence, even when the player runs out of budget she can still finish the game;
the worst case time is $d_0$. 
We assume the player has chosen a \emph{strategy} $\St$ that for each combination $(d,B)$ dictates
which distribution to choose in Step~1.
Of course, the idea is that this strategy is chosen so as to keep the expected number of rounds until $d=0$ is reached 
as small as possible.%
\pwtodo{I don't understand what the last two sentences are doing here. Are they supposed to come somewhere else? In the lower bound section?}
\mdtodo{In Thm. 1 the notion of a strategy is assumed to be known. It's here where it is explained.}

We give an example for a strategy. Say the initial budget is $B_0=1$.  
In distance $d$ the player bets $1/\log(d_0)$ on distance $\lfloor d/2 \rfloor$ and nothing on other distances. 
Since the expected waiting time for a jump with $i=\lfloor d/2 \rfloor$ is $1/\log(d_0)$, 
and since after $\log(d_0)$ such jumps the remaining distance $d$ cannot be larger than 2, 
the expected number of rounds is smaller than $\log^2(d_0)+2$.

\subsection{The main theorem}\label{subsect:game:theorem}

We assume that $B_0\ge1$. (This will be justified by the applications.)
The following theorem is the first main result of this paper.

\begin{theorem}\label{thm:budget_game}
  Assume the budget game is started with distance $d_0$ and a starting budget $B_0\ge1$,
  using an arbitrary fixed strategy $\St$.   
  Then the expected number of rounds is $\Omega\bigl(\log^2(d_0)/B_0\bigr)$.
\end{theorem}


For $1\le d \le n$ and $B\ge1$ let $T_{d,B}$ denote the expected number of rounds
until state 0 is reached when the budget game is run with strategy $\St$, starting in distance $d$ with budget $B$.
Theorem~\ref{thm:budget_game} states that $T_{d_0,B_0} = \Omega\bigl(\log^2(d_0)/B_0\bigr)$.

For proving the theorem, we proceed by induction.
The following lemma formulates the inductive assertion.
It is convenient to use the natural logarithm, here.
\begin{lemma}[Main Lemma]
  There is a constant $\alpha>0$ such that $T_{d,B} \ge \alpha\cdot\sln{d}/(1+B)$, for all $d\ge1$ and all $B\ge0$. 
\end{lemma}
Due to the assumption $B_0\geq 1$ in Theorem~\ref{thm:budget_game}, the theorem is immediately implied by the Main Lemma.
The offset by 1 in the denominator of the lower bound in the induction hypothesis (which does not occur in the corresponding lower bound of Theorem~\ref{thm:budget_game}) is necessary. 
In fact, since the budget game always finishes after $O(d_0)$ rounds, the theorem would not hold if we allowed very small values of $B_0$, \eg, for $B_0=o(\ln^2(d_0)/d_0)$.

The lemma is true for $d=1$, as $T_{1,B} = 1 > 0$.
By choosing $\alpha$ appropriately we may even say that it is
true for $1\le d \le d_{\text{base}}$ for any fixed $d_{\text{base}}$.   
\mdtodo{added: orientation}The induction step is involved: all of the following subsection is devoted to it.
The reader interested mainly in the application to the routing lower bound 
can safely skip this proof on first reading 
and continue with Sect.~\ref{sect:greedy:routing}.

\subsection{The inductive step}\label{subsect:induction:step}
It is obvious that the expected number of steps of the budget game for an optimal strategy
and the lower bound stated in the Main Lemma are monotone decreasing in $B$. 
On the other hand, it is not clear (and we will never use) that the expected number of steps of the game for an optimal strategy
is monotone in $d$.

Consider a distance $d\geq 2$ with remaining budget $B\geq 0$.
Let $L=\lfloor\log d\rfloor$.

Suppose strategy $\St$ dictates that in state $(d,B)$ the player chooses the valid bet $\beta$.
Recall that if step size $i\in\{2,\dots,d-1\}$ is chosen, the budget $B$ decreases by 
\begin{math}
  (i/d)^{3/2}\beta\bigl((i/2,d]\bigr),
\end{math}
and the remaining distance is $d-i$.
Thus, by the induction hypothesis, the remaining time is at least 
\begin{displaymath}
  \frac{\alpha\cdot\sln{d-i}}{1+B-(i/d)^{3/2}\beta\bigl((i/2,d]\bigr)}\ \text{, for $2\leq i<d$}.
\end{displaymath}
If step size $i=1$ is chosen, the budget remains $B$ and the remaining distance is $i-1$.
For $i=d$, the remaining time is 0.
Hence, the inductive assertion for $d$ follows from the following lemma:
\begin{lemma}\label{lem:budget-game-inductive-step}\samepage
  There is a constant $\alpha>0$ such that for any budget $B\geq 0$, distance $d\geq 2$, 
  and bet $\beta$ valid in state $(d,B)$ we have
  \begin{displaymath}
    1+\beta(1)\cdot\frac{\alpha\cdot\sln{d-1}}{1+B}
    \\+
    \sum_{2\leq i< d}\beta(i)\cdot\frac{\alpha\cdot\sln{d-i}}{1+B-(i/d)^{3/2}\beta\bigl((i/2,d]\bigr)}
    \\ \geq 
    \frac{\alpha\cdot\sln{d}}{1+B}.
  \end{displaymath}
\end{lemma}
The term 1 on the left hand side corresponds to the step that is being made in state $(d,B)$ and leads to a new state $d-i$.
The other terms in the sum are the lower bounds (as asserted by the induction hypothesis) 
for the remaining number of steps for each step size $i$, each multiplied with the probability that a steps of size $i$ is made.
There is no term for $i=d$, as after a step of size $d$ the game ends.

In the remainder of this subsection we prove Lemma~\ref{lem:budget-game-inductive-step}.

First note that for all $d\geq 2$ we have $\sln{d}-\sln{d-1}<1$.
Since $\beta(1)\leq 1$ and $B\geq 0$, we obtain
\begin{displaymath}
  \frac{1}{2}+\beta(1)\frac{\alpha\cdot\sln{d-1}}{1+B}\geq \beta(1)\frac{\alpha\cdot\sln{d}}{1+B}\quad
  \text{for $\alpha\leq 1/2$}.
\end{displaymath}
Thus, to prove Lemma~\ref{lem:budget-game-inductive-step} it suffices to show that
\begin{multline}\label{eq:10}
  \frac{1}{2}+\sum_{2\leq i< d}\beta(i)\cdot\frac{\alpha\cdot\sln{d-i}}{1+B-(i/d)^{3/2}\cdot\beta\bigl((i/2,d]\bigr)}
  \\ \geq
  \bigl(1-\beta(1)\bigr)\frac{\alpha\cdot\sln{d}}{1+B}\ ,
  \quad\text{for sufficiently small $\alpha>0$}.
\end{multline}
If $\beta(1)=1$, the right hand side evaluates to 0, so the claim is true.
Thus, in the following we assume $\beta(1)<1$.


We merge step sizes into orders of magnitude and consider the resulting probabilities, $g_\ell$, that a step size is of the order of magnitude $\ell$:
\begin{align*}
  g_\ell &:=\beta\left(\Bigl(\frac{d}{2^{\ell+1}}, \frac{d}{2^\ell}\Bigr]\right) \text{, for $\ell\in\{0,\dots,L-1\}$}.
\end{align*}
(Note that $d/2^L=d/2^{\lfloor\log d\rfloor}\in [1,2)$, and so $g_{L-1}=\beta\bigl([2,d/2^{L-2}]\bigr)$.)

Suppose step size $i\in\{2,\dots,d-1\}$ is applied.
Then for $\ell=\lfloor\log(d/i)\rfloor$, we have $i/2=d/2^{\log(d/i)+1}\leq d/2^{\ell+1}$ 
and $i/d=1/2^{\log(d/i)}\geq 1/2^{\ell+1}$.
Hence, we can bound the budget loss from below as follows (note that many terms are omitted!):
\begin{equation}\label{eq:budget_loss_bound}
  (i/d)^{3/2}\cdot\beta\bigl((i/2,d]\bigr)
  \geq
  \frac{g_\ell}{2^{3(\ell+1)/2}}.
\end{equation}
The new distance is
\begin{equation}\label{eq:25}
  d-i\geq d-d/2^\ell.
\end{equation}
Note that $\sln{z(1-x)}\geq \sln{z}-3x(\ln z)$ for all $z\ge1$ 
and all $x\in[0,\frac12]$ \fullversiononly{(see Claim~\ref{clm:square-ln-product} in the appendix)}\eabstronly{(see the full version of the paper)}.
Thus, for $\ell\ge1$, 
\begin{equation}\label{eq:30}
  \sln{d-d/2^\ell}
  =\sln{d(1-1/2^\ell)}
  \geq 
  \sln{d}-3(\ln d)/2^\ell
  .
\end{equation}
Hence, for each $\ell\in\{1,\dots,L-1\}$ we have 
\begin{multline}\label{eq:lower_bound_group_l}
  \sum_{d/2^{\ell+1}<i\leq d/2^\ell}\beta(i)\cdot\frac{\alpha\cdot\sln{d-i}}{1+B-(i/d)^{3/2}\cdot\beta\bigl((i/2,d]\bigr)}
  \stackrel{(\ref{eq:budget_loss_bound})}{\geq}
  \sum_{d/2^{\ell+1}<i\leq d/2^\ell}\beta(i)\cdot\frac{\alpha\cdot\sln{d-d/2^\ell}}{1+B-g_\ell/2^{3(\ell+1)/2}}
  \\ =
  g_\ell\cdot\frac{\alpha\cdot\sln{d-d/2^\ell}}{1+B-g_\ell/2^{3(\ell+1)/2})}.
  \stackrel{(\ref{eq:30})}{\geq}
  \alpha\cdot g_\ell\cdot (\ln d)\cdot\frac{\ln d-3/2^\ell}{1+B-g_\ell/2^{3(\ell+1)/2}}.
\end{multline}

For $\ell=0$ we need a stronger bound on the expectation of $\ln^2(d-i)$, 
which relies on the bound on the probability of obtaining large step sizes imposed by (B2).
We derive such a bound using the following claim.
\begin{claim}\label{clm:summation:close:to:target}
  Let $d^*=d - \lfloor d/2 \rfloor$ and let $\mu$ be an arbitrary probability distribution on $\{0,\ldots,d^*-1\}$
  with the property that $\mu(\{0,\ldots,j-1\}) \le 2j/d^*$, for $1\le j \le d^*$. 
  Then 
  \begin{equation}\label{eq:S100}
  \sum_{1\le i < d^*} \mu(i)\ln^2(i) \ge \ln^2(d) - O(\ln d).
  \end{equation}
\end{claim}
\begin{pproof}
  Since $\ln^2(i)$ is increasing in $i$, 
  the sum is minimal if $\mu$ puts as much weight as possible on the smallest numbers in $\{0,\ldots,d^*-1\}$. 
  By the inequality $\mu(\{0,\ldots,j-1\}) \le 2j/d^*$ it is clear that the distribution
  $$
  \mu(i)=\left\{
  \begin{array}{ll}
  {\displaystyle\frac{2}{{\textstyle d^*}}} & \text{, if }{\displaystyle0\le i < (d^*-1)/2}, \rule[-10pt]{0pt}{5pt}\\
  {\displaystyle\frac{1}{{\textstyle d^*}}} & \text{, if }{\displaystyle d^*\text{ is odd and } i = (d^*-1)/2}, \rule[-10pt]{0pt}{5pt}\\
  0 & \text{, if }{\displaystyle(d^*-1)/2 < i < d^*},
  \end{array}
  \right.
  $$
  will achieve this minimum. 
  Thus
  \begin{equation}\label{eq:S200}
  \sum_{1\le i < d^*} \mu(i)\ln^2(i) \ge  \frac{2}{d^*}\cdot\sum_{1\le i < (d^*-1)/2}\!\!\ln^2(i).
  \end{equation}
  We assume that $d\ge3$ and that $d^*$ is even 
  (and hence $i < d^*/2$ if and only if $i < (d^*-1)/2$); the calculation for odd $d^*$ is similar.
  Utilizing that $\ln^2(x)$ is increasing in $x$ we may easily bound the last sum from below by comparing it with an integral,
  as follows.
  \begin{eqnarray*}
  \lefteqn{\sum_{1\le i < d^*/2} \!\!\ln^2(i) 
  \ge 
  \int_1^{d^*/2}\!\!\ln^2(x)\,dx - \ln^2(d^*/2)} 
  \\ &=& \left[x\ln^2(x) - 2x\ln x +2x\right]_1^{d^*/2} - 2\ln^2(d^*/2)
  \\ &=& (d^*/2)\ln^2(d^*/2) - d^*\ln (d^*/2) + (d^* -2 -2\ln^2(d^*/2))
  \\ &\ge& (d^*/2)\ln^2(d^*/2) - d^*\ln (d^*/2).
  \end{eqnarray*}
  In combination with (\ref{eq:S200}) we obtain
  $$
  \sum_{1\le i < d^*} \mu(i)\ln^2(i) >   \ln^2(d^*/2) - 2\ln (d^*/2).
  $$
  Now $d^*/2 \ge d/4$, and hence 
  $\ln^2(d^*/2) - 2\ln (d^*/2) \ge \ln^2(d/4) - 2\ln (d/4) > \sln{d} - 2(\ln4)(\ln d) -2\ln d= \ln^2(d) - O(\ln d)$.
Thus Claim~\ref{clm:summation:close:to:target} is proved.\qed
\end{pproof}

Let $d^*=d - \lfloor d/2 \rfloor$. 
Recall that $g_0=\beta\bigl((d/2,d]\bigr)=\beta\bigl([\lfloor d/2 \rfloor+1,d]\bigr)=\beta\bigl([d-d^*+1,d]\bigr)$.
We define a probability distribution $\mu$ on  $\{0,\ldots,d^*-1\}$ by $\mu(i)= \beta(d-i)/g_0$. 
According to property (B2) $\mu$ satisfies 
\begin{displaymath}
  \mu\bigl([0,j-1]\bigr)=\frac{\beta\bigl([d-j+1,d]\bigr)}{g_0} \le \frac{2j}{d^*}
  \quad \text{for $1\leq j\leq d^*$.}
\end{displaymath}
Hence we can apply Claim~\ref{clm:summation:close:to:target} to see that there is a constant $C'$ (the same one for all $d$, $B$, and all valid budget bets $\beta$), such that 
\begin{multline}\label{eq:lower_bound_group_0}
  g_0\cdot(\ln d)(\ln d - C')
  \leq
  g_0\cdot\sum_{1\leq i < d^*} \mu(i)\cdot\ln^2(i)
  =
  \sum_{d-d^\ast<j\leq d-1} g_0\cdot\mu(d-j)\cdot\ln^2(d-j)  
  \\ =
  \sum_{\lfloor d/2\rfloor < j < d} \beta(j)\cdot\ln^2(d-j)
  =
  \sum_{d/2<j < d}\beta(j)\cdot\ln^2(d-j).
\end{multline}
Let $C:=\max\{3,C'\}$.
We bound the sum on the left hand side of (\ref{eq:10}) from below by grouping terms according to order of magnitude of the corresponding step sizes and by applying (\ref{eq:lower_bound_group_0}) for group $\ell=0$ and (\ref{eq:lower_bound_group_l}) for all other groups.
(In order to bound the denominator for group $\ell=0$ we use relation (\ref{eq:budget_loss_bound}).)
\begin{multline}\label{eq:60}
  \sum_{2\leq i< d}\beta(i)\cdot\frac{\alpha\cdot\sln{d-i}}{1+B-(i/d)^{3/2}\cdot\beta\bigl((i/2,d]\bigr)}
  \\ \geq
  \alpha\cdot g_0\cdot(\ln d)\cdot\frac{\ln d-C'}{1+B-g_0/2^{3/2}}
  +
  \sum_{1\leq \ell\leq L-1}\alpha\cdot g_\ell\cdot(\ln d)\cdot\frac{\ln d-3/2^\ell}{1+B-g_\ell/2^{3(\ell+1)/2}}
  \\ \geq
  \alpha\cdot (\ln d)\cdot \sum_{0\leq \ell\leq L-1}g_\ell\cdot\frac{\ln d-C/2^\ell}{1+B-g_\ell/2^{3(\ell+1)/2}}
\end{multline}

Applying this to (\ref{eq:10}), we see that to prove Lemma~\ref{lem:budget-game-inductive-step} it suffices to show
\begin{multline}\label{eq:100}
  Z:=
  \frac12+\alpha\cdot (\ln d)\cdot \sum_{0\leq \ell< L}g_\ell\cdot\frac{\ln d-C/2^\ell}{1+B-g_\ell/2^{3(\ell+1)/2}}
  -
  \bigl(1-\beta(1)\bigr)\frac{\alpha\cdot\sln{d}}{1+B} 
  \geq 0.
\end{multline}

Let $\gamma_\ell=g_\ell/\bigl(1-\beta(1)\bigr)$ for $\ell=0,\dots,L-1$ and let $y=(1+B)/\bigl(1-\beta(1)\bigr)$.
(Recall that we need only consider the case $\beta(1)<1$.)
Then $y\ge1$ and $g_\ell=\gamma_\ell\cdot(1+B)/y$, and so it suffices to show that $Z'\geq 0$, where
\begin{multline}\label{eq:110}
  Z'
  :=
  Z\cdot\frac{y}{\alpha\cdot\ln d}
  =
  \frac{y}{2\alpha\cdot\ln d}
  +
  \sum_{0\leq\ell< L}\gamma_\ell\cdot(1+B)\cdot\frac{\ln d-C/2^\ell}{1+B-\gamma_\ell(1+B)/(y\cdot 2^{3(\ell+1)/2})}
  -\ln d
  \\ =
  \frac{y}{2\alpha\cdot\ln d}
  +
  \sum_{0\leq\ell< L}\gamma_\ell\cdot\frac{\ln d-C/2^\ell}{1-\gamma_\ell/(2^{3(\ell+1)/2} y)}
  -\ln d.
\end{multline}

Recall that 
\begin{displaymath}
  \sum_{0\leq\ell<L}g_\ell=\beta\bigl([2,d]\bigr)=1-\beta(1).
\end{displaymath}
Hence $\sum_{0\leq\ell<L}\gamma_\ell=1$, and so
\begin{equation}\label{eq:145}
  \ln d=\sum_{0\leq\ell<L}\gamma_\ell\cdot\ln d.
\end{equation}
Note also that
\begin{equation}\label{eq:150}
  1+\frac{\gamma_\ell}{2^{3(\ell+1)/2}y}\leq 3/2\quad \text{for all $0\leq\ell<L$},
\end{equation}
because $y\geq 1$ and $\gamma_\ell\leq 1$.
Moreover, $1/(1-x)\geq 1+x$ for all real numbers $x<1$, and using this inequality for $x=\gamma_\ell/(2^{3(\ell+1)/2}y)$ yields
\begin{multline}\label{eq:before:cauchy:schwarz}
  Z'
  \stackrel{(\ref{eq:145})}{\geq}
  \frac{y}{2\alpha\cdot\ln d}
  +
  \sum_{0\leq\ell< L}\gamma_\ell
    \left(
      \left(\ln d-\frac{C}{2^\ell}\right)\left(1+\frac{\gamma_\ell}{2^{3(\ell+1)/2}y}\right)
      -\ln d
    \right)
  \\ =
  \frac{y}{2\alpha\cdot\ln d}
  +    
  \sum_{0\leq\ell< L}\gamma_\ell
  \left(
    \frac{\gamma_\ell \ln d}{2^{3(\ell+1)/2}y}-\frac{C}{2^\ell}\left(1+\frac{\gamma_\ell}{2^{3(\ell+1)/2}y}\right)
  \right)
  \\ \stackrel{(\ref{eq:150})}{\geq}
  \frac{y}{2\alpha\cdot\ln d}
  +
  \frac{\ln d}{2^{3/2}y}\sum_{0\leq\ell< L}\frac{{\gamma_\ell}^2}{2^{3\ell/2}}-\frac{3C}{2}\sum_{0\leq\ell< L}\frac{\gamma_\ell}{2^{\ell}}.
\end{multline}
A lower bound for the second summand
can be obtained using the following inequality,
which is proved using the Cauchy-Schwarz inequality.

\begin{claim}\label{clm:cauchy:schwarz:application}
  Let $x_1,\dots,x_m\in\IR$.
  Then 
  \begin{displaymath}
     \Bigl(\sum_{1\leq i\leq m}\frac{x_i}{2^i}\Bigr)^2
     \leq
     (2+\sqrt2)\cdot\sum_{1\leq i\leq m}\frac{{x_i}^2}{2^{3i/2}}.
  \end{displaymath}
\end{claim}
\begin{pproof}
  Recall that the Cauchy-Schwarz inequality states
  \begin{displaymath}
    \Bigl(\sum_{1\leq i\leq m} a_ib_i\Bigr)^2
    \leq
    \Bigl(\sum_{1\leq i\leq m} {a_i}^2\Bigr)\cdot\Bigl(\sum_{1\leq i\leq m}{b_i}^2\Bigr)
  \end{displaymath}
  for arbitrary reals $a_1,b_1,\dots,a_m,b_m$.
  Define $a_i=x_i/2^{3i/4}$ and $b_i=1/2^{i/4}$.
  Then
  \begin{multline*}
     \Bigl(\sum_{1\leq i\leq m}\frac{x_i}{2^i}\Bigr)^2
     =
     \Bigl(\sum_{1\leq i\leq m} a_ib_i\Bigr)^2
    \leq
    \Bigl(\sum_{1\leq i\leq m} {a_i}^2\Bigr)\cdot\Bigl(\sum_{1\leq i\leq m}{b_i}^2\Bigr)
     =
    \Bigl(\sum_{1\leq i\leq m} \frac{{x_i}^2}{2^{3i/2}}\Bigr)\Bigl(\sum_{1\leq i\leq m} \frac{1}{2^{i/2}}\Bigr).
  \end{multline*}
  The claim now follows immediately from the geometric series
  \begin{displaymath}
    \sum_{1\leq i\leq m}{\frac{1}{2^{i/2}}}
    < 
    \sum_{i\ge1}{\frac{1}{2^{i/2}}}
    =
    \frac{1/\sqrt{2}}{1-1/\sqrt2}=2+\sqrt{2}.
  \end{displaymath}\qed
\end{pproof}

Applying Claim~\ref{clm:cauchy:schwarz:application} to the sum in the second term of (\ref{eq:before:cauchy:schwarz})
we obtain
\begin{displaymath}
  \sum_{0\leq \ell<L}\frac{{\gamma_\ell}^2}{2^{3\ell/2}}\geq \left(\sum_{0\leq\ell<L}\gamma_\ell/2^{\ell}\right)^2\cdot\frac{1}{2+\sqrt{2}}.
\end{displaymath}
Hence, defining $S=\sum_{0\leq\ell< L}\gamma_\ell/2^\ell$ and using that ${2^{3/2}\cdot(2+\sqrt2)}=9.6568\ldots<10$, we obtain
\begin{multline}\label{eq:160}
  Z'
  \geq 
  \frac{y}{2\alpha\cdot\ln d}
  +
  \frac{\ln d}{2^{3/2}(2+\sqrt 2)y}\left(\sum_{0\leq\ell< L}\frac{\gamma_\ell}{2^{\ell}}\right)^2
  -
  \frac{3C}{2}\sum_{0\leq\ell< L}\frac{\gamma_\ell}{2^{\ell}}
  \\ >
  \frac{y}{2\alpha\cdot\ln d}
  +
  \frac{\ln d}{10y}\cdot S^2
  -
  \frac{3C}{2}\cdot S.
\end{multline}
Clearly, neither $y/(2\alpha\ln d)$ nor $\ln d/(10y)$ are negative.
If $y\geq 3CS\alpha\ln d$, then $y/(2\alpha\ln d)\geq 3CS/2$ and so $Z'\geq 0$.
Thus, it suffices to consider $y<3CS\alpha\ln d$.
In this case
\begin{displaymath}
  Z'
  \geq
  \frac{\ln d}{30CS\alpha\cdot\ln d}\cdot S^2
  -
  \frac{3C}{2}S    
  =
  S\cdot C\cdot\frac{3}{2}\left(
    \frac{1}{\alpha\cdot 45\cdot C^2}
    -
    1
  \right).
\end{displaymath}
This term is non-negative for $\alpha\leq 1/\left(45\cdot C^2\right)$.
This completes the proof of Lemma~\ref{lem:budget-game-inductive-step}, 
and thus the induction step in the proof of Theorem~\ref{thm:budget_game}.
\qed

\section{Application to Greedy Routing in Small-World Graphs}
\label{sect:greedy:routing}

\subsection{Greedy routing on a grid}\label{subsect:greedy routing}
Consider the $D$-dimensional infinite grid $\ZD$, for $D\ge2$.
The distance measure used is the ``Manhattan distance'' induced by the ``Manhattan norm''
$$
\|(x_1,\ldots,x_D)\|_1 = |x_1|+\cdots+|x_D|.
$$
Fix some distribution $\lambda$ over $\ZD$, which we will call the \emph{long range contact (short: lrc) distance distribution}.
Every node $u$ on $\ZD$ 
has a \emph{long range contact} that is determined by the following random experiment 
(executed independently for all nodes $u$):
First, a distance $r$ is picked according to the lrc distance distribution $\lambda$, 
and then $u$'s long range contact is chosen uniformly at random among all nodes $v$ with $\mdist{u-v}=r$.
(The combination of these experiments defines the augmenting distribution.)
We consider the directed random graph in which every node in $\ZD$ has one outgoing edge 
to each of its neighbors on the grid, and one edge to its long range contact.

We are interested in the expected greedy-routing time from an arbitrarily fixed source node, $s$, to a fixed target node, $t$.
The greedy-routing time is the length of the path from $s$ to $t$ 
that is obtained by following in each step the edge that minimizes the Manhattan distance to $t$ 
(ties broken arbitrarily).
For symmetry reasons we may assume w.l.o.g.\ that $t=0^D=(0,\dots,0)$.
Our main result is summarized in the following theorem.
\begin{theorem}\label{thm:grid-routing}
Let $\lambda$ be an arbitrary lrc distance distribution.
Assume $\ZD$ is augmented with long range contacts chosen according to $\lambda$.   
 Then for every starting node $s\in \ZD$ the expected greedy routing time from $s$ to $0^D$ is $\Omega\left(\log^2\mdist{s}\right)$.
\end{theorem}

\begin{remark}\upshape 
For simplicity we formulate the theorem in terms of the infinite grid,
but it applies equally well to the $D$-dimensional finite grid
$G_D=\{0,\ldots,n-1\}^D$ with edge length $n$ (with the edges induced from $\ZD$ 
and the Manhattan distance measure)
and the $D$-dimensional finite torus $T_D=\{0,\ldots,n-1\}^D$
(with the edges of $G_D$ plus ``wraparound edges'' from 
$(x_1,\ldots,n-1,\ldots,x_D)$ to $(x_1,\ldots,0,\ldots,x_D)$,
and the shortest path distance).
The statements carry over directly as long as $s$ and $t$ have distance smaller than $n/4$
(on $T_D$) and have distance at least $2\mdist{s-t}$ from the border of the grid (on $G_D$).  
\end{remark}

The rest of this section shows 
how to utilize Theorem~\ref{thm:budget_game} to prove Theorem~\ref{thm:grid-routing}.

\subsection{Simulation of routing by a budget game}\label{subsect:simulation}
To prove Theorem~\ref{thm:grid-routing}, we wish to apply 
our results for the budget game to greedy routing. 
As an intermediate step, we introduce a game $G$,
played in rounds on $\ZD$, which generalizes greedy routing. 
The state of game $G$ is a distance $d\in\IN$; the initial state is some $d_0$.
When in state $d>0$, the player chooses a point $u\in\ZD$ with $\mdist{u}=d$ arbitrarily.
Then, using the lrc distribution $\lambda$ as described above, 
she carries out a random experiment to obtain a long range contact $v$ of $u$.
The round ends with the new state being $d'=\min\{\mdist{v},d-1\}$.
The game ends once state $d=0$ has been reached.
The cost of a run of $G$ is the number of rounds carried out. 

A \emph{strategy} for the player is a function that
for each distance $d$ gives some vertex $u$ with $\mdist{u}=d$.
If the player adheres to a fixed strategy for choosing $u$ in each step,
the cost becomes a random variable that only depends on the 
random experiments used for finding the long range contacts. 
An \emph{optimal} strategy is one that minimizes the expected cost.
(For given $\lambda$, 
an optimal strategy can be determined by induction on $d=0,1,2,\ldots$.)

It is obvious that if game $G$ is played starting from $d_0$ with an optimal strategy, 
then the expected cost (\ie, the expected number of rounds until the game ends) 
is not larger than the expected greedy routing time in $\ZD$ 
from any start node at distance $d_0$ to $0^D$. 
Thus, it suffices to prove that for any start distance $d_0$ 
and any strategy the expected number of rounds of game $G$ 
is $\Omega\bigl((\log d_0)^2\bigr)$.

Now assume an arbitrary strategy $\St_G$ for game $G$ has been fixed.
Then for every distance $d$
strategy $\St_G$ induces a probability distribution $\beta_d$ on $\{1,\dots,d\}$ as follows:
$\St_G$ dictates which node $u$ with $\mdist{u}=d$ to choose.
Node $u$ has some grid neighbors of distance $d-1$ from $0^D$ and one long range contact $v$,
chosen at random on the basis of lrc distribution $\lambda$, as described before.
One chooses some neighbor $v$ of $u$ that minimizes $\mdist{v}$, 
and advances by $d-\mdist{v}$ (${}\ge1$) to the new state $\mdist{v}$.
For $1\le i \le d$, $\beta_d(i)$ is defined as the probability that the progress
$d - \min\{d-1,\mdist{v}\}$ is equal to $i$. 

Now we can describe how, given a strategy $\St_G$ for game $G$,
the game $G$ played with $\St_G$ may be simulated
by a budget game $G_{\text{B}}$ played with a strategy $\St_{\text{B}}$.
The initial budget is some constant $B_0$ (independent of $\St_G$; value to be determined later).
The initial distance is $d_0$, as in game $G$.
When at distance $d$, the bet by the player of the budget game 
is the probability distribution $\beta_d$ on $\{1,\ldots,d\}$ as defined
in the previous paragraph.
The effect is that in one round of $G_{\text{B}}$ 
the player moves from state $d$ to state $d-i$ with probability $\beta_d(i)$, for $1\le i \le d$,  
exactly as in game $G$ with strategy $\St_G$.
From this it is obvious that the expected number of rounds game $G_{\text{B}}$ needs to finish 
is exactly the same as in game $G$ with strategy $\St_G$.

Note, however, that before we can say that $G_{\text{B}}$ is a budget game in the technical
sense, so that the results of Section~\ref{sect:budget:game}
can be applied, we have to make sure that the defining 
constraints on the probability distributions $\beta_d$ used in the bets of $G_{\text{B}}$ are satisfied.  
In a round in which one moves from state $d$ to state $d-i$, $i\geq 2$, 
the budget shrinks by $(i/d)^{3/2}\beta_d\bigl((i/2,d]\bigr)$. 
(For $i=1$ the budget remains unchanged.)
For $G_{\text{B}}$ to be a genuine budget game
it is sufficient that $\beta_d$ is always a valid budget bet, 
\ie, that constraints (B1) and (B2) on the probability distribution $\beta_d$ are satisfied
when state $d$ is reached with remaining budget $B$ in the budget game.
If this is the case, then the simulation $G_{\text{B}}$ of game $G$ with strategy $\St_G$
is a legitimate budget game, and Theorem~\ref{thm:budget_game} can be applied.

In the following, we develop two lemmas which show that $\beta_d$ always satisfies (B1) and (B2) 
if the constant starting budget $B_0$ is large enough.
This will then prove Theorem~\ref{thm:grid-routing}.

\subsection{The budget suffices}\label{subsect:budget:suffices}
 
The first lemma deals with property (B1);
it implies that if in $G_{\text{B}}$ state $d$ is reached and the remaining budget is $B$, 
then $B$ is large enough to accommodate the next bet.


\begin{lemma}\label{lem:network-game-sufficient-budget}\samepage
  Let $d_k\in[1,d_0)$ be a state that can be reached in game $G$ after $k$ rounds, \ie, there are states
  $d_k<d_{k-1}<\dots<d_1<d_0$ and step sizes $i_s=d_s-d_{s+1}$, $0\leq s<k$, such that $\beta_{d_s}(i_s)>0$.
  Then for $B_0\geq 21$,
  \begin{equation}\label{eq:network-game-sufficient-budget}
    B_0-\sum_{\substack{0\leq s<k\\i_s>1}}\left(\frac{i_s}{d_s}\right)^{3/2}\beta_{d_s}\bigl((i_s/2,d_s]\bigr)\geq 1.
  \end{equation}
\end{lemma}
Note that in order to satisfy (B1) when at distance $d_k$, 
it would even suffice that the remaining budget is $\beta_{d_k}\bigl([2,d_k]\bigr)$, 
while our lower bound on the remaining budget is 1.

\begin{pproof}[of Lemma~\ref{lem:network-game-sufficient-budget}]
Suppose a node $u$ with $\mdist{u}=d_u$ has a long range contact $v$ with $\mdist{v}=d_v<d_u$.
By the triangle inequality, $\mdist{v-u}\geq d_u-d_v$, and since $d_v<d_u$, we have $\mdist{v-u}\leq d_v+d_u<2d_u$.
Thus, in game $G$ in order to go from state $d$ to some state in $[0,d']$, $d'\leq d-2$, 
the node $u$ with $\mdist{u}=d$ 
(which is chosen by the player) 
must have chosen a long range contact $v$ such that $d-d'\leq \mdist{v-u} < 2d$.
It follows that 
\begin{equation}\label{eq:beta-bounded-by-lambda}
  \beta_{d}\bigl([d-d',d]\bigr)\leq \lambda\bigl([d-d',2d)\bigr)\ \text{for any $1\leq d'\leq d-2$}. 
\end{equation}
We define $\lambda(I)=\sum_{i\in I}\lambda(i)$ for any set $I\subseteq\IN$.
It suffices to show that
\begin{displaymath}
  B_0-\sum_{\substack{0\leq s<k\\i_s>1}}\left(\frac{i_s}{d_s}\right)^{3/2}\lambda\bigl((i_s/2,2d_s)\bigr)\geq 1.
\end{displaymath}
We have
\begin{multline*}
  \sum_{\substack{0\leq s<k\\i_s>1}}\left(\frac{i_s}{d_s}\right)^{3/2}\lambda\bigl((i_s/2,2d_s)\bigr)
  =
  \sum_{\substack{0\leq s<k\\i_s>1}}\left(\frac{i_s}{d_s}\right)^{3/2}\left(\sum_{i_s/2<j< 2d_s}\lambda(j)\right)
  \\ =
  \sum_{2\leq j< 2d_0}\lambda(j)\cdot\left(\sum_{\substack{0\leq s < k: \\ 1\le i_s/2<j< 2d_s}}\left(\frac{i_s}{d_s}\right)^{3/2}\right).
\end{multline*}
Since $\sum_{2\leq j<2d_0}\lambda(j)\leq 1$, it suffices to show for every $j\in[2,2d_0)$ that
\begin{equation}\label{eq:sufficient-budget-to-show}
    B_0-1\geq \sum_{\substack{0\leq s< k: \\  1 \le i_s/2<j< 2d_s}}\left(\frac{i_s}{d_s}\right)^{3/2}.
\end{equation}
Fix $j\in[2,2d_0)$ arbitrarily.
For $\ell\in\bigl[0,\lfloor\log d\rfloor\bigr]$ let $S(\ell)$ be the set of indices $s\in\{0,\dots,k-1\}$ such that 
\begin{displaymath}
  1\le i_s/2<j< 2d_s\ \text{and}\ d_s\in(d_0/2^{\ell+1},d_0/2^\ell].
\end{displaymath}
Then
\begin{equation}\label{eq:sufficient-budget-c}
  c_\ell
  :=
  \sum_{s\in S(\ell)}\left(\frac{i_s}{d_s}\right)^{3/2}
  \leq
  \left(\frac{2^{\ell+1}}{d_0}\right)^{3/2}\sum_{s\in S(\ell)}{i_s}^{3/2}.
\end{equation}
Let $s_0$ be the smallest index in $S(\ell)$, \ie, state $d_{s_0}$ is reached in the budget game before any of the step sizes $i_s$, $s\in S(\ell)$, are applied.
Since all step sizes $i_s$, $s\in S(\ell)$, can still be subtracted from $d_{s_0}$ without reaching state 0, we have
$\sum_{s\in S(\ell)}i_s< d_{s_0}\leq d_0/2^\ell$.
In addition, $i_s< 2j$ for all $s\in S(\ell)$, so by convexity of the function $x\mapsto x^{3/2}$,
\begin{displaymath}
  \sum_{s\in S(\ell)} {i_s}^{3/2}
  < 
  \frac{d_0/2^\ell}{2j}\cdot (2j)^{3/2}
  =
  \frac{d_0}{2^\ell}\cdot (2j)^{1/2}.
\end{displaymath}
Using this in (\ref{eq:sufficient-budget-c}) yields
\begin{displaymath}
  c_\ell\leq 4\cdot\left(j\cdot\frac{2^\ell}{d_0}\right)^{1/2}.
\end{displaymath}
Now note that $S(\ell)=\emptyset$ and thus $c_\ell=0$ for $d_0/2^{\ell}\leq j/2$, \ie, for $\ell\geq \log(d_0/j)+1$.
Hence,
\begin{multline*}
    \sum_{\substack{0\leq s\leq k: \\ 1 \le i_s/2<j< 2d_s}}\left(\frac{i_s}{d_s}\right)^{3/2}
    =
    \sum_{0\leq\ell\leq\log d}c_\ell
    \leq
    \sum_{0\leq\ell<\log(d_0/j)+1}4\cdot\left(j\cdot\frac{2^\ell}{d_0}\right)^{1/2}
    \\ \leq
    4\cdot\left(\frac{j}{d_0}\right)^{1/2}\cdot\sum_{0\leq\ell<\log(d_0/j)+1}2^{\ell/2}
    <
    4\cdot\left(\frac{j}{d_0}\right)^{1/2}\cdot\frac{(\sqrt{2})^{\log(d_0/j)+2}-1}{\sqrt{2}-1}    
    \\ \leq
    4\cdot\left(\frac{j}{d_0}\right)^{1/2}\cdot\frac{2\cdot(d_0/j)^{1/2}}{\sqrt{2}-1}
    <
    20.
\end{multline*}
This proves (\ref{lem:network-game-sufficient-budget}) for $B_0\geq 21$, and thus the lemma.
\qed
\end{pproof}

\subsection{A geometry lemma}\label{subsect:geometry}

It remains to show that the probability distributions $\beta_d$
that arise in the simulation of game $G$ by $G_{\text{B}}$  
satisfy condition (B2). 
As before, let $d^*=d-\lfloor d/2 \rfloor$;
recall that $(d/2,d] = [d-d^*+1,d]$, a set of size $d^*$.  
\begin{lemma}\label{lem:network-game-geometry}
  For all distances $d\ge3$ and all $1 \le j\leq d^*$,
  \begin{displaymath}
   \beta_d\bigl([d-j+1,d]\bigr)\leq\frac{2j}{d^*}\cdot\beta_d([d-d^*+1,d]).
  \end{displaymath}
\end{lemma}

For $w\in\ZD$ the the Manhattan ball with radius $t$ around $w$,
which is $\{x\in\ZD \mid \|x-w\|_1\le t\}$,  
is denoted by $B_{w,t}$, and the Manhattan sphere with radius $t$ around $w$,
which is $\{x\in\ZD \mid \|x-w\|_1 = t\}$, is denoted by $S_{w,t}$. 
 
Recall how $\beta_d$ is defined.
The player in game $G$ chooses a node $u$ with $\mdist{u}=d$.
The lrc distribution $\lambda$ is used to choose some distance $r$;
afterwards the long range contact $v$ itself is chosen
uniformly at random from $S_{u,r}$. 
It is obvious that it is sufficient to prove the inequality in Lemma~\ref{lem:network-game-geometry}
under the condition that $u$ and $r$ are fixed.  
Note that the jump distance $i = d-\mdist{v}$ is in $[d-j+1,d]$ if and only if
$\mdist{v} \le j-1$, which means that $v$ is in the Hamming ball $B_{0,j-1}$.
The jump distance is in $[d-d^*+1,d]$ if and only if $\mdist{v} \le d^*-1$.
We can restrict out attention to lrc lengths $r$ with which one can reach at least one such point $v$.     
Because on $S_{u,r}$ the uniform distribution is used,
for proving Lemma~\ref{lem:network-game-geometry} it is sufficient to 
show the following. 

\begin{lemma}\label{lem:geometry}
Let $u\in\ZD$ with $\|u\| = d$, and let $d^\ast= d - \lfloor d/2 \rfloor$. 
Assume that $S_{u,r}$ intersects $B_{0,d^\ast-1}$. 
Then for $1\le j \le d^\ast$ the following holds:
$$
\frac{|B_{0,j-1} \cap S_{u,r}|}{|B_{0,d^\ast-1} \cap S_{u,r}|} \le \frac{2j}{d^\ast}.
$$
\end{lemma}
In words: If $j\le d^\ast$, then among the set of points on a sphere with radius $r$ around $u$
that are strictly closer to the origin than $d^\ast$
not more than a fraction of $2j/d^\ast$ is strictly closer to the origin 
than $j$.

\textbf{Note.} Assume $D=2$. With $u$ on the $x_1$-axis one can achieve that 
$2j-1$ points are in $B_{0,j-1} \cap S_{u,r}$ and $d^\ast+j$ points are in $B_{0,d^\ast-1} \cap S_{u,r}$.
Assuming $d^\ast$ is large and $j$ is small shows that the factor $2$ in the lemma is best possible. 

\emph{Proof}.\quad
We start with some preliminary remarks.  
From $S_{u,r} \cap B_{0,d^\ast-1}\neq\emptyset$ it follows that 
$d \le r + d^\ast-1  = r + d- \lfloor d/2 \rfloor -1$, hence $r \ge \lfloor d/2 \rfloor +1 \ge d^\ast$. 
If $B_{0,j-1} \cap S_{u,r}=\emptyset$, there is nothing to show. 
So, if needed, we will assume that $B_{0,j-1} \cap S_{u,r}\neq\emptyset$.

A \emph{vertex} (of $S_{u,r}$) is an element of $S_{u,r}$ that maximizes or minimizes
one of the components. There are $2D$ vertices;
they have the form $u\pm (0,\ldots,0,r,0,\ldots,0)$, with exactly one nonzero component
in the second summand. 
It is easy to see that $B_{0,d^\ast-1}$ can contain at most one of these vertices:
two vertices have Manhattan distance exactly $2r$, 
but two points in $B_{0,d^\ast-1}$ cannot be further apart than $2(d^\ast-1)$, which is smaller than $2r$.

If $B_{0,d^\ast-1}$ contains one vertex $w$, 
the intersection $B_{0,d^\ast-1} \cap S_{u,r}$ does not change if $u$ is moved away from $w$
along the line that connects $w$ and $u$, and $r$ is increased accordingly. 
The same is true if  $B_{0,d^\ast-1}$ does not contain a vertex at all. 
We see that the statement we need to prove does not really depend on $r$ and $u$,
only on the shape of the intersection $B_{0,d^\ast-1} \cap S_{u,r}$.
A moment's thought shows that is sufficient to show the following (and apply it for $d^\ast=i$).   

\emph{Claim}: 
Consider two balls $B_{0,i-1}$ and $B_{0,j-1}$ around the origin, 
with $1\le j \le i$.
Assume that some sphere $S_{u,r}$ intersects $B_{0,i-1}$ in such a way that $B_{0,i-1}$ 
contains at most one vertex of $S_{u,r}$. 
Then
$$
Q = Q(i,j,u,r) = \frac{|B_{0,j-1} \cap S_{u,r}|}{|B_{0,i-1} \cap S_{u,r}|} \le \frac{2j}{i}.
$$
The claim is proved by induction on $D$. 

\textbf{Basis:} $D=2$. 

For symmetry reasons we may assume that the leftmost (``westernmost'') vertex of $S_{u,r}$ is in $B_{0,i-1}$, if any. 
We may further assume that $r$ is large enough so that certain manipulations
with $S_{u,r}$ are possible without another vertex coming close to $B_{0,i-1}$.
We may assume 
that $B_{0,j-1} \cap S_{u,r}\neq\emptyset$ and  $2j<i$, 
since otherwise the claim is trivially true. We consider some cases.

\emph{Case} 1: $B_{0,i-1}$ does not contain a vertex of $S_{u,r}$. ---
Then the intersection $B_{0,i-1} \cap S_{u,r}$ is a straight line with $i$ or $i-1$ points.
The intersection $B_{0,j-1} \cap S_{u,r}$ is either empty (and hence $Q=0$) or 
a straight line with $j$ or $j-1$ points. Hence $Q \le j/(i-1) \le (j+1)/i \le 2j/i$.
\begin{figure}[h!]
\begin{center}
\includegraphics[width=0.28\textwidth]{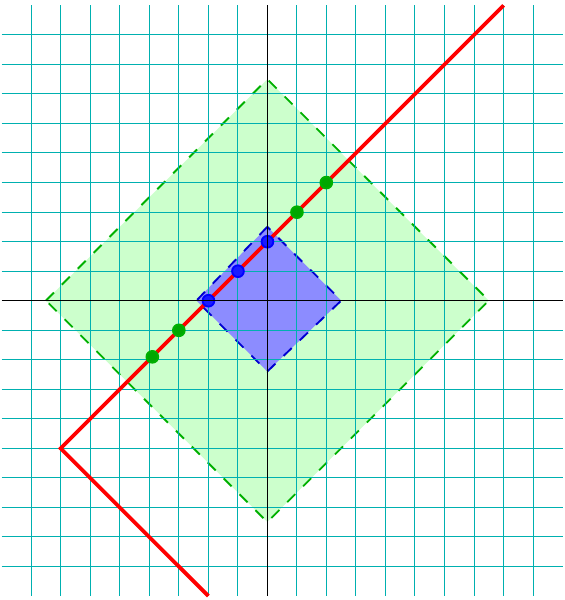}
\end{center}
\caption{Case 1. $B_{0,i-1}$ (large ball) is green/lighter, $B_{0,j-1}$ (small ball) is blue/darker, 
the bent line is a part of $S_{u,r}$ (red). 
$|B_{0,i-1} \cap S_{u,r}|=7=i-1$ and $|B_{0,j-1} \cap S_{u,r}|=3=j$.}
\label{fig:geometry:case:one}
\end{figure}

\emph{Case} 2. $B_{0,i-1}$ contains the leftmost vertex of $S_{u,r}$, but this vertex is not in $B_{0,j-1}$. ---
Then the intersection $B_{0,j-1} \cap S_{u,r}$ is
a straight line with $j$ or $j-1$ points. We move $u$ parallel to this line until the leftmost vertex of $S_{u,r}$
moves out of $B_{0,i-1}$. This change decreases 
the denominator $|B_{0,i-1} \cap S_{u,r}|$, but leaves the numerator $|B_{0,j-1} \cap S_{u,r}|$ unchanged. 
Applying Case 1 yields the desired inequality.
\begin{figure}[h!]
\begin{center}
\includegraphics[width=0.28\textwidth]{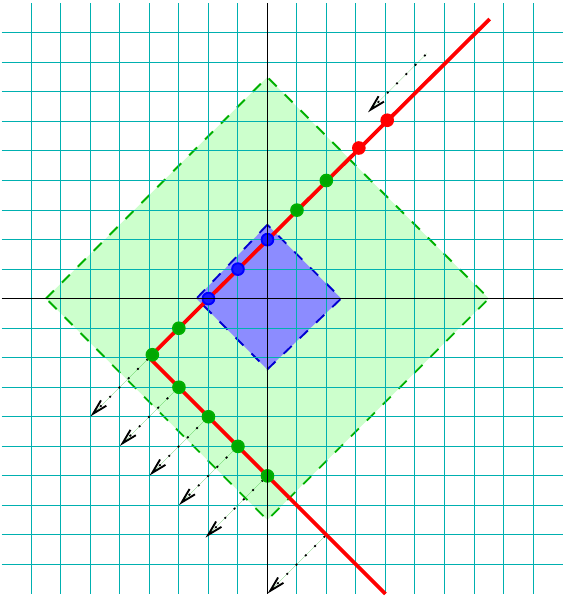}
\end{center}
\caption{Case 2. $B_{0,i-1}$, $B_{0,j-1}$, and $S_{u,r}$ as in Fig.~\ref{fig:geometry:case:one}. 
Moving $S_{u,r}$ to the southwest until the vertex is outside $B_{0,j-1}$ does not change $S_{u,r}\cap B_{0,j-1}$ 
and makes the intersection with $S_{u,r} \cap B_{0,i-1}$ smaller.}
\label{fig:geometry:case:two}
\end{figure}   

\emph{Case} 3: The leftmost vertex of $S_{u,r}$ is a point $(z,0)$ on the $x_1$-axis, inside $B_{0,j-1}$.
Then $-j < z < j$. Let $y=z-1+j$, hence $0\le y \le 2j-2$. 
Then $|B_{0,j-1} \cap S_{u,r}|$ is either $2j-y-2 = j-z-1$ or 1 more than this; in any case we have $|B_{0,j-1} \cap S_{u,r}| \le j-z$.
Similarly, $|B_{0,i-1} \cap S_{u,r}| \ge i-z-1$. Hence $Q \le (j-z)/(i-z-1)$, which is a number $<1$ (since $2j<i$, hence $j<i-1$).
Increasing numerator and denominator by the same summand increases the fraction, hence
$$
Q\le \frac{j-z}{i-z-1} \le \frac{j-z}{i-z-1} \le \frac{j + j}{i + j -1} \le \frac{2j}{i}.
$$
\begin{figure}[h!]
\begin{center}
\includegraphics[width=0.28\textwidth]{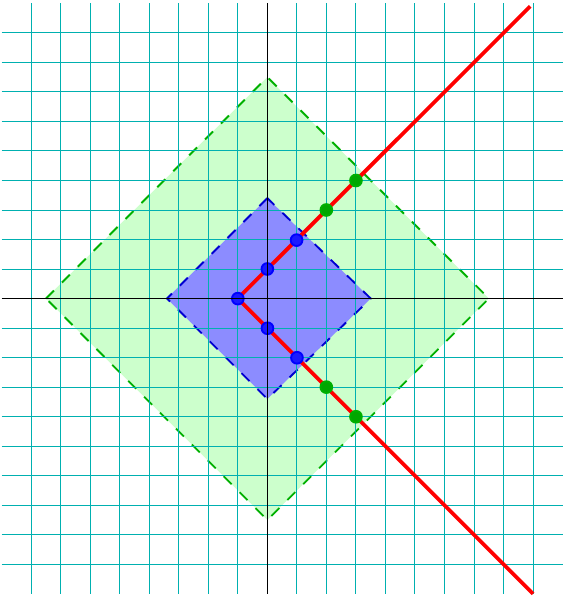}
\end{center}
\caption{Case 3. $B_{0,i-1}$, $B_{0,j-1}$, and $S_{u,r}$ as in Fig.~\ref{fig:geometry:case:one}. 
Here: $i=8$, $j=4$.
The leftmost vertex of $S_{u,r}$ is $(-1,0)=(z,0)$. Hence $y=2$.
$|B_{0,j-1} \cap S_{u,r}|=j - z=5$, while $|B_{0,i-1} \cap S_{u,r}|=i - z=9$.
We have $5/9 < 8/8 =2j/i$.}
\end{figure} 
  
\emph{Case} 4: The leftmost vertex $(z,z')$ of $S_{u,r}$ is inside $B_{0,j-1}$, but not on the $x_1$-axis. 
Assume $z'>0$, so it is above the $x_1$ axis. Move $u$ in southwest direction until it hits the $x_1$-axis. 
This means that the leftmost vertex moves to $(z-z',0)$. These changes increase both $|B_{0,j-1} \cap S_{u,r}|$
and $|B_{0,i-1} \cap S_{u,r}|$ by $z'$, hence $Q$ increases. Applying Case 3 yields the claim also in this case.

\begin{figure}[h!]
\begin{center}
\includegraphics[width=0.28\textwidth]{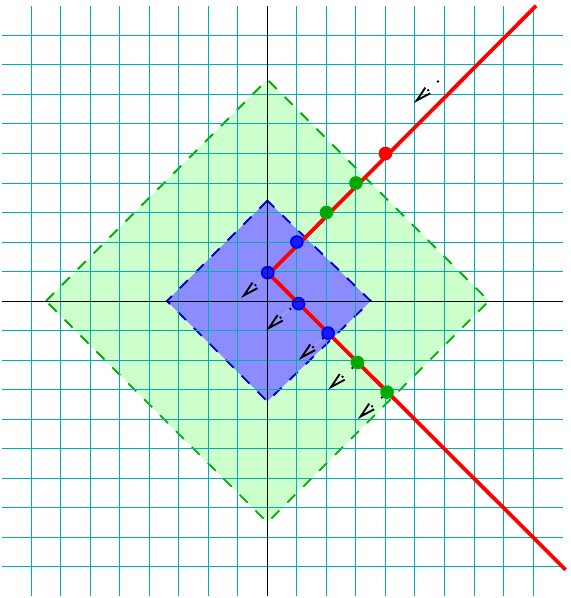}
\end{center}\caption{Case 4. $B_{0,i-1}$, $B_{0,j-1}$, and $S_{u,r}$ as in Fig.~\ref{fig:geometry:case:one}. 
Moving $S_{u,r}$ in southwest direction until the leftmost vertex hits the $x_1$-axis
increases $|B_{0,j-1} \cap S_{u,r}|$ (by 1) and does not change $|(B_{0,i-1} - B_{0,j-1}) \cap S_{u,r}|$.}
\label{fig:geometry:case:four}
\end{figure}    

\textbf{Induction step:} $D > 2$.

As induction hypothesis we assume that the claim is true for $D-1$. 
We cut $\ZD$ into hyperplanes (``layers'')
$L_s=\{(s,x_2,\ldots,x_D) \mid x_2,\ldots,x_D \in\IZ\}$, for $s\in\IZ$.
If $j-|s|\le0$, the intersection $B_{0,j-1} \cap S_{u,r} \cap L_s$ of our configuration with layer $L_s$ is empty,
so it does not contribute to $|B_{0,j-1} \cap S_{u,r}|$, and we may ignore this layer. 
Assume $ |s| < j $. 
Intersecting $L_s$ with $B_{0,i-1}$,  $B_{0,j-1}$, and $S_{u,r}$, and ignoring dimension 1 gives
 balls with radius $i-|s|-1$ and $j-|s|-1$ around the origin and a sphere $S_{u',r'}$,
all in $\IZ^{D-1}$. The crucial observation is that only one vertex of $S_{u',r'}$ can lie in $B_{0,i-1}$.
The induction hypothesis yields that restricted to $L_s$ the ratio between points in $B_{0,j-1}$ and $B_{0,i-1}$
is $Q(i-|s|,j-|s|,u',r')  \le 2(j-|s|)/(i-|s|) \le 2j/i$. 
Summing this over all layers $L_s$, for $-j < s <j$, yields that 
$Q(i,j,u,r) \le \frac{2j}{i}$. 

This ends the induction step, and the proof of the claim, and the proof of the lemma.
\qed

\section{Conclusion}
We presented the first lower bound for greedy routing in augmented $D$-dimensional grids and tori for $D\geq 2$, for any uniform and isotropic augmenting distribution.
Together with the tight lower bound for rings and lines~\cite{DW2009a}, this settles the question whether for all dimensions $D$ the inverse $D$-th power distribution is the optimal uniform and isotropic augmenting distribution for greedy routing in Kleinberg's small world graph model.
It seems to us that the new lower bound technique presented here is easier to understand than the one applied in \cite{DW2009a} for the one-dimensional case.
The analysis of budget games yields interesting insights into the reasons why faster greedy routing is not possible.
We also think that budget games are an interesting concept in their own right.

We believe that it is not hard to extend our lower bound technique to prove a lower bound for the greedy diameter of the line (\ie, to obtain the same result as in \cite{DRWW2010a}). 
To obtain a tight lower bound for greedy routing on the ring between two randomly chosen nodes as in \cite{DW2009a} may not be as easy, but we conjecture that it is possible.

Our lower bounds hold regardless whether every node has one or $k$ independently chosen long range contacts, as long as $k$ is a constant.
For super-constant $k$ this is not clear.
For $k=O(\log n)$ and any constant $D\geq 1$, greedy routing on the $D$-dimensional torus or grid augmented with the inverse $D$-th power distribution takes only $O\bigl((\log n)^2/k\bigr)$ time.
It seems likely that our lower bound technique can be generalized somehow to show that this is optimal for all uniform and isotropic augmenting distributions, but a full proof seems to require significant additional effort.

\bibliographystyle{plain}
\bibliography{literature}

\fullversiononly{
\newpage
\appendix

\section{Additional Proof}\label{sect:app}
%
%
%
\begin{claim}\label{clm:square-ln-product}
  For all $z\ge1$ and $x\in[0,\frac12]$ we have
  \begin{displaymath}
    \ln(z(1-x))^2\geq \ln^2(z)-3x(\ln z).
  \end{displaymath} 
\end{claim}
\begin{pproof}
 The claim is obvious for $z=1$, so we may assume $z>1$. Note that $\ln(1-x)\geq -3x/2$ for all $x\in[0,\frac12]$.
  This is immediate for the boundary cases $x=0$ and $x=1/2$ and thus follows for all $x\in(0,\frac12)$ 
  since the function $x\mapsto \ln(1-x)$ is concave.
  Hence,
\begin{displaymath}
    \ln^2(z(1-x))
    =
    \ln^2 z+2(\ln z)(\ln (1-x))+\ln^2(1-x)
     \geq
    \ln^2 z-3x(\ln z). 
\end{displaymath}\qed
\end{pproof}
}

\end{document}